\documentclass[superscriptaddress,twocolumn,showpacs,prl]{revtex4-1}

\usepackage{graphicx}
\usepackage{amsmath, amsthm, amssymb}

\begin{document}

\title{Dynamically Correlated Region in Sheared Colloidal Glasses Revealed by Neutron Scattering}
\author{Zhe Wang}
\affiliation{Department of Engineering Physics, Tsinghua University, Beijing 100084, People's Republic of China}
\affiliation{Biology and Soft Matter Division, Oak Ridge National Laboratory, Oak Ridge, TN 37831, USA}
\author{Takuya Iwashita}
\affiliation{Department of Electrical and Electronic Engineering, Oita University, Oita 870-1192, Japan}
\author{Lionel Porcar}
\affiliation{Institut Laue-Langevin, B.P. 156, F-38042 Grenoble CEDEX 9, France}
\author{Yangyang Wang}
\affiliation{Center for Nanophase Materials Sciences, Oak Ridge National Laboratory, Oak Ridge, TN 37831, USA}
\author{Yun Liu}
\affiliation{Center for Neutron Research, National Institute of Standards and Technology, Gaithersburg, MD 20899-6100, USA}
\author{Luis E. S\'anchez-D\'iaz}
\affiliation{Biology and Soft Matter Division, Oak Ridge National Laboratory, Oak Ridge, TN 37831, USA}
\author{Bin Wu}
\affiliation{Biology and Soft Matter Division, Oak Ridge National Laboratory, Oak Ridge, TN 37831, USA}
\author{Takeshi Egami}
\affiliation{Department of Materials Science and Engineering and Department of Physics and Astronomy, The University of Tennessee, Knoxville, TN 37996-1508, USA}
\author{Wei-Ren Chen}
\email[Corresponding author: ]{chenw@ornl.gov}
\affiliation{Biology and Soft Matter Division, Oak Ridge National Laboratory, Oak Ridge, TN 37831, USA}

\begin{abstract}
The microscopic deformation mechanism of charged colloidal glasses with extended-range interactions under shear is investigated by in-situ small-angle neutron scattering, and a dynamically correlated region (DCR) is identified.  This short-lived region provides the resistance to the configurational rearrangement imposed by the external deformation, as evidenced by the evolution of the size of DCR in the shear thinning regime and the quantitative agreement between the local stress sustained by DCR and the macroscopic stress from rheological measurements at low and mediate shear rates. This finding suggests that DCR is an important quantity for microscopically addressing the flow and deformation behavior of strongly interacting colloids.
\end{abstract}

\date{\today}

\maketitle
In recent years there has been increasing interest in understanding the fundamental mechanism that controls the deformation behavior of disordered materials \cite{bt1}. One reason for the current excitement stems from the microscopic description of flow based on the ``dynamic heterogeneity'' - the spatial inhomogeneity in the relaxation dynamics \cite{bt1,bt2}. For example, computational study has been carried out to understand how the local configurational rearrangements and their cooperative organization influence the plasticity of amorphous solids \cite{barrat1,lem1,lem2,lem3,lem4,barrat2,lin1}. In highly supercooled liquids, the shear thinning phenomenon was found by simulations to be the consequence of diminishing inhomogeneity of flow due to the increasingly frequent configurational fluctuations \cite{yama1,yama2,yama3,tanaka1}.
 
There also exists an extensive amount of studies on the effects of local plastic rearrangement in developing a microscopic description of flow \cite{bt1,weeks1,poon1}. From the deviation of affine deformation condition, results of confocal microscopy experiments on sheared supercooled colloidal liquids and glasses have identified the plastic component of deformation \cite{weeks2,weeks3,poon2}. The spatial correlation of localized plastic arrangements was examined \cite{schall1,schall2,schall3,schall4,schall5} and its connection to the non-monotonic flow curves, namely the shear banding instability, was further investigated \cite{schall6}.
   
As a deeper understanding of the deformation behavior of amorphous materials is gained, the role played by inter-particle interactions demands a full investigation.  While its profound influence on the rheology of amorphous materials has been recognized \cite{russel1,wagner1,oswald1,coussot1}, the microscopic mechanism through which the increasing cooperativity promoted by extended-range interaction influences the deformation behavior remains to be elucidated. The motivation of this study is to identify this process from the perspective of dynamical heterogeneity. 

In this Letter, we investigate the influence of the inter-particle potential on the flow characteristics of concentrated colloidal suspensions using rheometry and small angle neutron scattering (SANS). By extending the spatial range of the inter-particle potential, a transition from a colloidal liquid to a colloidal glass is observed. The analysis of SANS spectra shows that the mechanical response of a colloidal glass to the imposed shear is localized in a \textit{dynamically correlated region} (DCR), which is promoted by the extended-range inter-particle potential. The correlation between the DCR and the mechanical behavior is supported by the agreement between the microscopic stress revealed by scattering and the macroscopic stress measured by rheometry.

We studied two representative colloidal suspensions at the same volume fraction of 0.4: a charged stabilized colloidal suspension and a hard-sphere suspension. In the former system the inter-particle potential is characterized by a extended-range electrostatic interaction \cite{dhont1} while in the latter system only the hard-core repulsion is present \cite{watanabe1}. The Kob-Andersen mixture \cite{kob1} of monodispersed particles of two different sizes (120 nm and 80 nm) was used to avoid crystallization. As demonstrated in Fig. 1, the rheological properties of the suspensions show a strong dependence on the inter-particle potential. In the linear viscoelastic regime shown in Fig. 1(a), the hard spheres exhibit a liquid-like behavior. In contrast, the dynamic moduli of the charged colloids indicate that the sample is an elastic solid in the quiescent state. Results of steady shear measurements given in Fig. 1(b) reveal that the charged colloids displays much higher viscosity and more dramatic shear thinning compared to the hard spheres at low and moderate shear rates. These observations provide an initial clue for understanding the prominent role of the extended-range potential in determining the rheological properties of colloidal suspensions. 

\begin{figure}[h]
\centering
\includegraphics[scale=0.75]{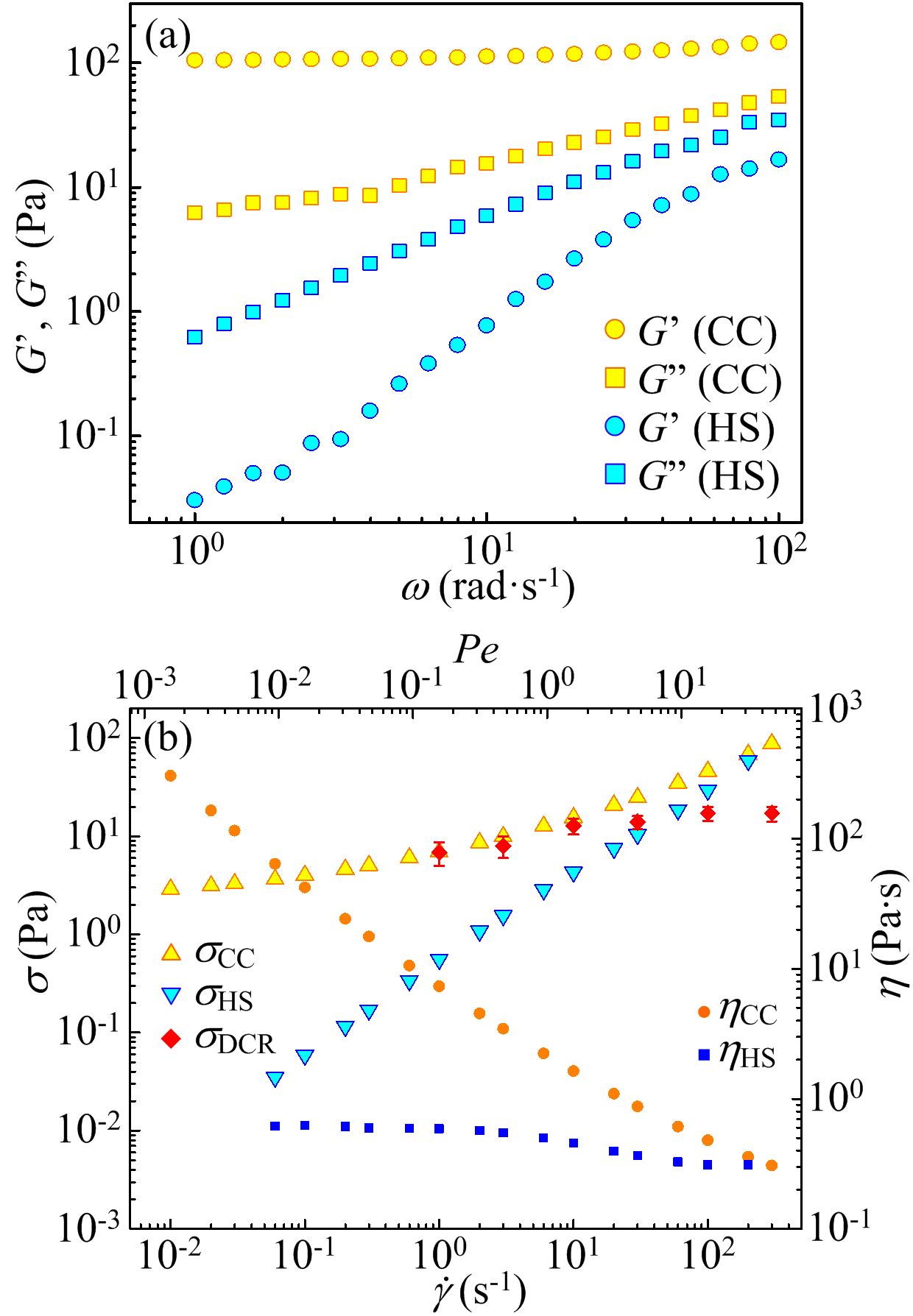}
\caption{Rheological behavior of the charged colloidal (CC) suspension and the hard-sphere (HS) suspension. (a) Frequency dependence of the storage and loss moduli $G^\prime$ and $G^{\prime\prime}$ for the two samples. (b) Shear stress $\sigma$ and viscosity $\eta$ as a function of $\dot\gamma$ and P\'eclet number (\textit{Pe}). Red rhombuses: shear stress sustained by the DCR ($\sigma_{DCR}$).}
\label{F1}
\end{figure}

We use SANS to investigate the microscopic origin of the observed nonlinear rheological behavior. Figure 2(b) and (c) show the SANS spectra obtained from both the flow-velocity gradient ($v-\nabla v$ or 1 - 2) and flow-vorticity ($v-\omega$ or 1 - 3) planes for the two samples. When subjected to steady shear, the scattering profiles present elliptical shapes in both configurations. In neither configuration no noticeable scattering signature of shear-induced ordering, such as layer formation, is observed within the probed range of shear rate $\dot\gamma$. A similar development is also observed by our complementary Brownian dynamics (BD) simulation \cite{sm}.  Trajectory analysis suggests the origin of intensity variation is the local ordering promoted by the anisotropic density fluctuation, instead of the long-range layering. 

\begin{figure}[h]
\centering
\includegraphics[scale=0.85]{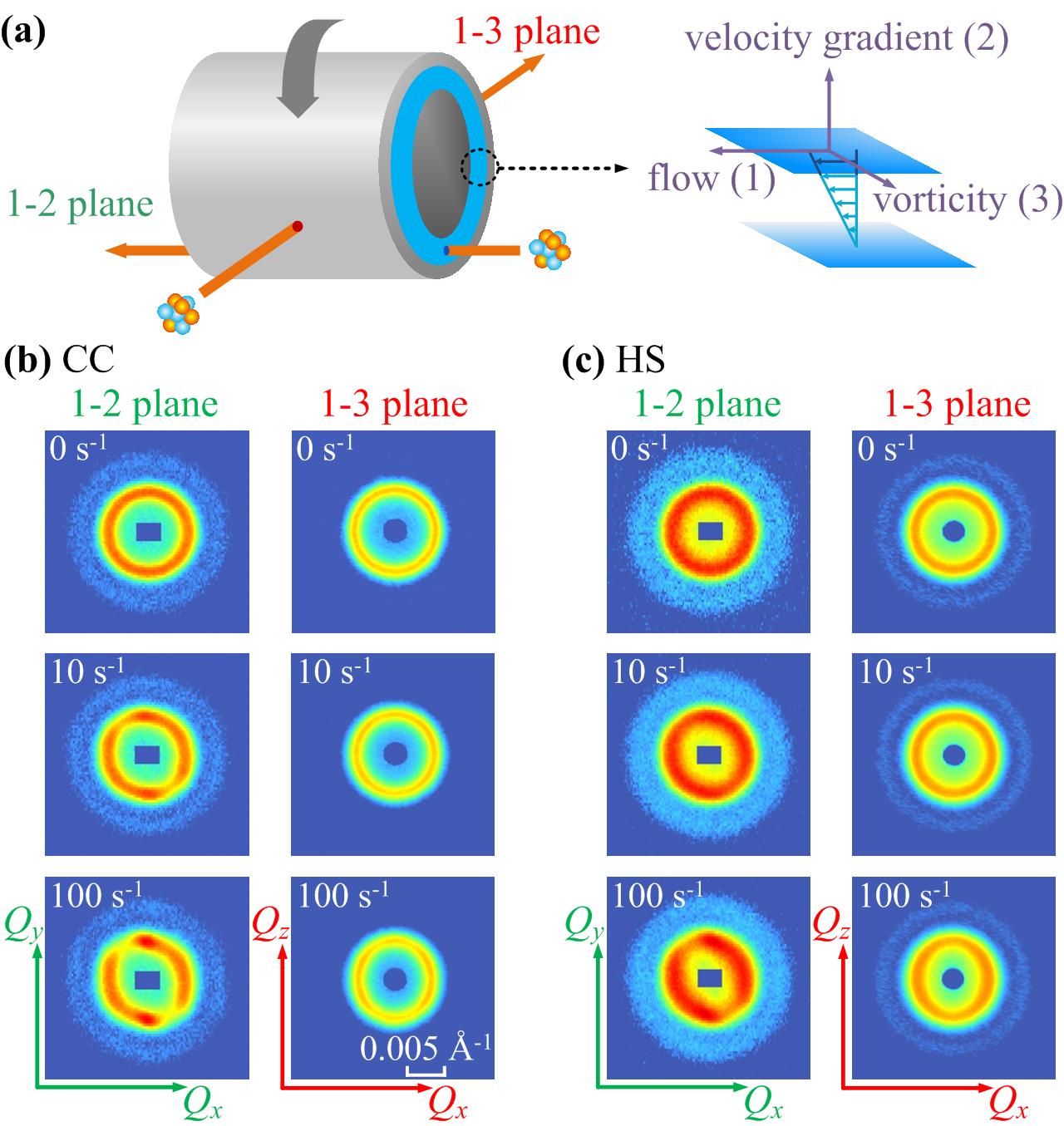}
\caption{(a) Illustrations of the Rheo-SANS experiment under Couette geometry. 1, 2 and 3 denote the directions of flow ($v$), velocity gradient ($\nabla v$) and vorticity ($\omega$), respectively. (b) and (c) 2D SANS spectra obtained from flow-velocity gradient (or 1-2) plane and from flow-vorticity (or 1-3) plane at $\dot\gamma$=0,10 and 100 s$^{-1}$ for charged colloidal (CC) suspension and hard-sphere (HS) suspension. }
\label{F2}
\end{figure}

To address the connection between the spatial correlation functions and the flow behavior of the system, we adopt a spherical harmonic expansion (SHE) approach for the SANS data analysis. The pair distribution function (PDF) $g(r)$ of a sheared fluid can be expressed explicitly by SHE as \cite{hess1,hess2,hess3,hess4,hess5,ackerson1,johnson1,schall7,wagner2,exp0}:
\begin{equation}
g(\vec r)=\sum_{l,m}g_l^m(r)Y_l^m\left( \vec r/r \right),
\end{equation}
where $Y_l^m\left( \frac{\vec r}{r}\right)$ are the real spherical harmonic functions and $g_l^m(r)$ the expansion coefficients. The harmonic functions are indexed by the order $l$ and the degree $m$. The relevant $g_l^m(r)$ can be determined from the anisotropic SANS intensity $I(\vec Q)$. More details are given in the Supplemental Material \cite{sm}. Due to the symmetry imposed by steady shear, $g_2^{-2}(r)$ has been recognized to be the most relevant coefficient which connects the shear-induced structural distortion to the macroscopic properties \cite{soloman1,rice1}. For an elastic solid undergoing an affine deformation, $g_2^{-2}(r)$ is proportional to the derivative of the quiescent PDF $g(r)$ when the shear strain $\gamma$ is sufficiently small \cite{egami1,exp1}. Namely,
\begin{equation}
g_2^{-2}(r)=-\frac{\gamma}{\sqrt{15}}r\frac{dg(r)}{dr}.
\end{equation}

The validity of Eq. 2 in sheared dense fluids was demonstrated by a computational study \cite{ashurst1}, which evidences that the concept of elasticity can be applied to the time-averaged local response of highly cooperative fluids under steady shear. In Fig. 3(a)-(d) we plot the comparisons of $g_2^{-2}(r)$ and $-rdg(r)/dr$ for charged colloids at $\dot\gamma$=3,10,30 and 100 $\mathrm{s}^{-1}$, respectively.  It is seen that the characteristic variations of these two functions are generally in phase, which qualitatively agrees with the prediction of Eq. 2. This observation suggests that the system is essentially elastically deformed at these shear rates, even the system is flowing. Such deformation coherency is also observed in a simulation study on a metallic liquid \cite{egami2} and our BD results \cite{sm}, but has not been reported in prior computational studies of nonlinear rheology of charged colloids \cite{grest1,wagner3,morris1}. The same analysis was applied to the hard spheres and the result is shown in Fig. 3(e) and (f). Though the corresponding SANS spectra are still featured by the anisotropy between the compressional and extensional directions as shown in Fig. 2, the oscillation of each $g_2^{-2}(r)$ lags behind that of the $-rdg(r)/dr$. The invalidity of Eq. 2 in the hard spheres evidences that the coherent elastic deformation no longer exists without the presence of the extended-range electrostatic repulsion. The deformation is dominated by plastic flows characterized by configurational rearrangement. These phenomena are also found in our complementary BD simulations \cite{sm}.

\begin{figure}[h]
\centering
\includegraphics[scale=0.9]{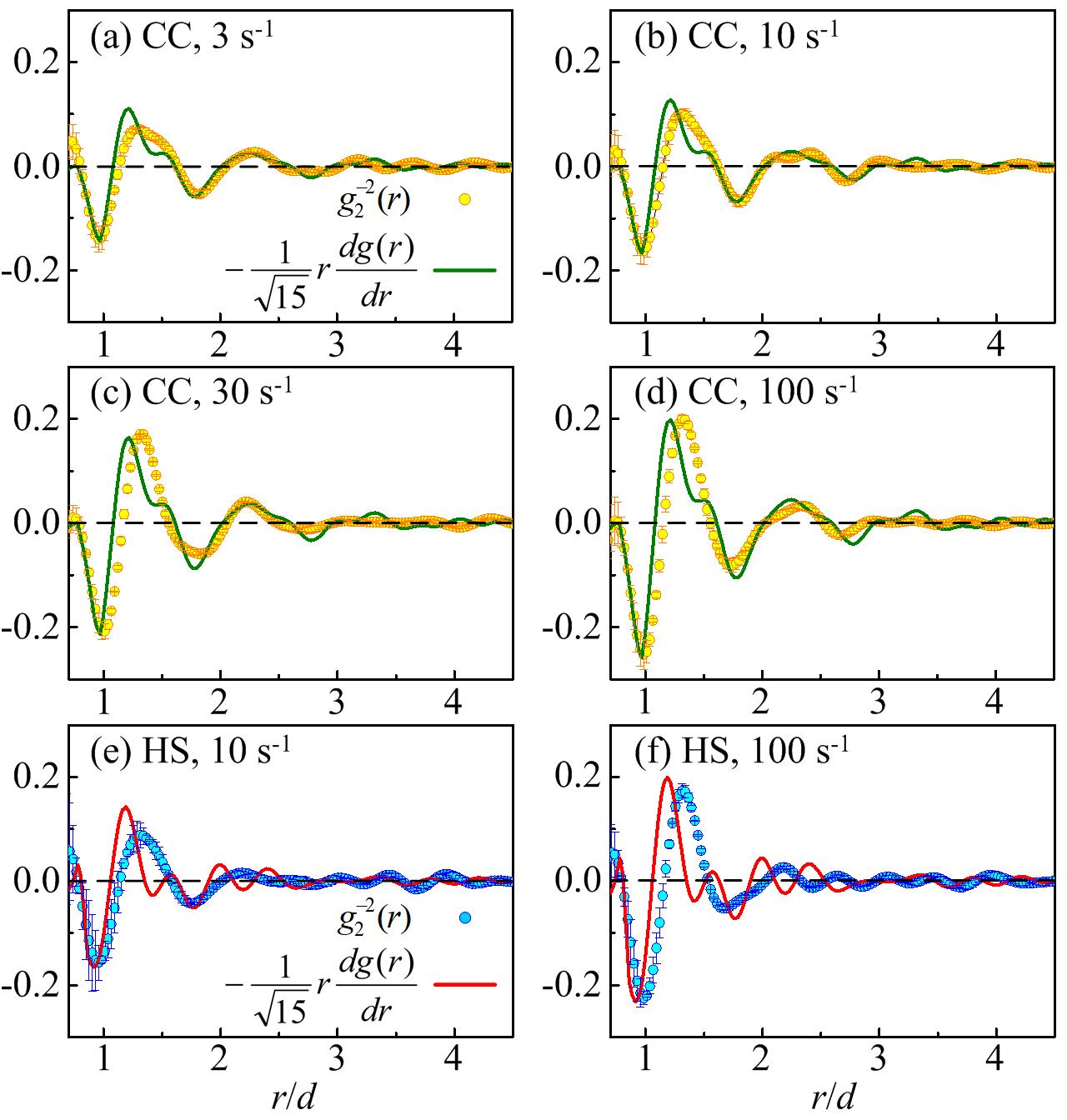}
\caption{Comparison between $g_2^{-2}(r)$ (circles) and $rdg(r)/dr$ (lines). (a)-(d) give the results for the charged colloidal suspension at $\dot\gamma$=3,10,30 and 100 s$^{-1}$, respectively. (e) and (f) are the results for the hard-sphere suspension at $\dot\gamma$=10 and 100 s$^{-1}$, respectively. The magnitude of $-1/\sqrt{15}rdg(r)/dr$ is scaled to match that of $g_2^{-2}(r)$ for all panels.}
\label{F3}
\end{figure}

In a random stacking of particles, the local configurational environment is known to differ from one tagged particle to another. As a result, the constant strain picture given by Eq. 2 does not provide a complete description about the microscopic deformation.  To further elucidate the structure of the flowing elasticity, we introduce the dependence of $\gamma$ on the spatial range over which the elastic deformation is sustained \cite{egami2}:
\begin{equation}
\gamma(r)\equiv -\sqrt{15}g_2^{-2}(r)/\left[ rdg(r)/dr \right].
\end{equation}

\begin{figure}[h]
\centering
\includegraphics[scale=0.9]{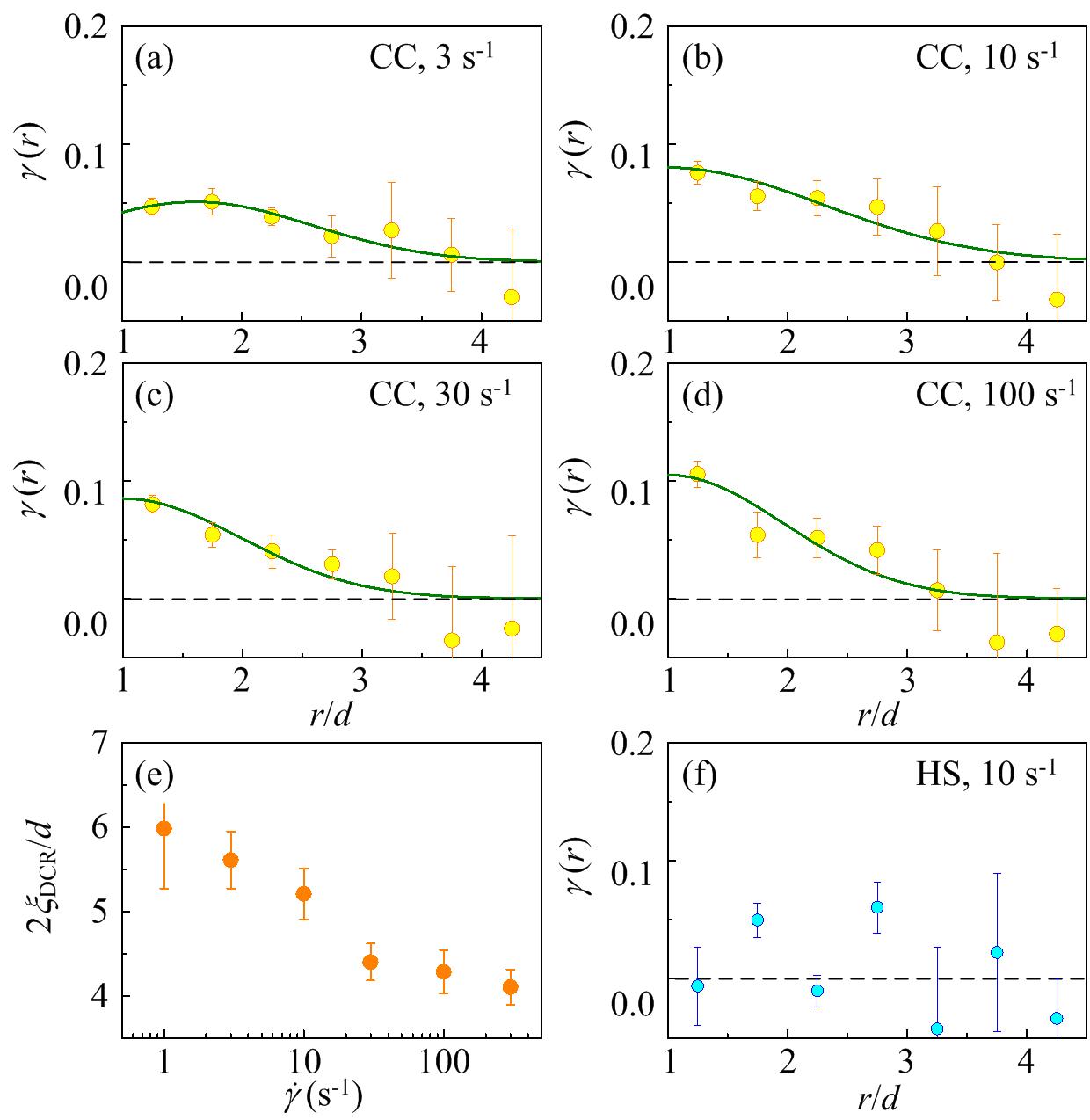}
\caption{(a)-(d): $\gamma(r)$ of the charged colloidal (CC) suspension determined by Eq. 3 at $\dot\gamma$=3,10,30 and 100 s$^{-1}$, respectively (yellow circles). The green curves denote the fitting results by the Gaussian function. (e) summarizes the size of DCR $2\xi_{DCR}$ as a function of $\dot\gamma$ ̇ for the CC suspension. (f) gives the result of the hard-sphere (HS) suspension at $\dot\gamma$=10 s$^{-1}$. No effective DCR is observed in this case.}
\label{F4}
\end{figure}

Figure 4(a) to (d) presents the $\gamma (r)$ for the charged colloids at $\dot\gamma$=3,10,30 and 100 s$^{-1}$. We would like to point out that the extraction of $\gamma (r)$ does not involve any model fitting but only Fourier or Bessel transforms and data binning. A region of nonzero $\gamma (r)$ with a spatial range of several particle diameter $d$ is observed for all measured $\dot\gamma$. We name this region \textit{dynamically correlated region} (DCR): This region is short-lived and dynamically fluctuating. Within its spatial range, the local structure undergoes an elastic deformation with an average strain given by $\gamma (r)$ when the system is under steady shear. Beyond this region, the particle motion is dominated by liquid-like random displacements. The existence of DCR suggests dynamical heterogeneity in the mechanical response of the system to applied shear. $\gamma (r)$ can be considered as a correlation function describing a cooperative region characterized by mechanical coherency in the flow. Based on the previous simulation study \cite{egami2}, a Gaussian function is used to model the landscape of $\gamma (r)$:
\begin{equation}
\gamma(r)\approx \gamma_M \mathrm{exp}\left[ -(r-p)^2/2\delta_{DCR}^2 \right],
\end{equation}
where $p$ is the peak position, $\delta_{DCR}$ is the standard deviation of the Gaussian distribution, and $\gamma_M$ is the average maximum strain of DCR. The fit curves with Eq. 4 are also shown in Fig. 4(a) to (d).  Accordingly, a length scale $\xi_{DCR}=p+\sqrt{2\mathrm{ln}2}\delta_{DCR}$ is defined to denote the correlation length of the cooperatively elastic deformation in the steady flow.  The size of the DCR, $2\xi_{DCR}$, is shown in Fig. 4(e). A decrease of elastic coherency is revealed by the decrease of the DCR size from about $6d$ to $4d$ as $\dot\gamma$ increases from 1 to 300 s$^{-1}$. Meanwhile, an increase in $\gamma_M$ from about 0.045 to 0.11 is found. The $\gamma (r)$ of the hard spheres at $\dot\gamma$=10 s$^{-1}$ is shown in Fig. 4(f). As expected, no discernible DCR is observed.

The above analysis reveals two different microscopic mechanisms in the flowing colloids.  In the charged colloids, the changes of the momentum and position of a particle can instantaneously influence surrounding particles through the extended-range electrostatic interaction. Consequently, the particles within a certain spatial range undergo elastic coherent deformation in response to the imposed shear.  During this process, a reference particle retains its original neighbors until the stress generated by shear is sufficient to cause local configurational rearrangement. The deformation and yielding of DCR are ubiquitous and persistently successive at the particle level. In contrast, the suspension of hard spheres exhibits a completely different microscopic picture: The motion of one particle influences others mainly through collisions due to the lack of the extended-range electrostatic ``restoring force''.  The exchange of momentum and energy between particles is no longer instantaneous instead retarded by the mean free time between collisions.  This causes the mismatch between $g_2^{-2} (r)$ and $-rdg(r)/dr$ as shown in Fig. 3(e) and (f), and leads to the breakdown of the local deformation coherency.  Therefore, we conclude that the extended-range electrostatic repulsion acts as an energy barrier to resist the applied strain, and the hindrance of topological rearrangement by DCR is the principle mechanism of the significant enhancement of viscosity in the charged colloids compared to the hard spheres at low and mediated shear rates, as demonstrated in Fig. 1(b).

Having established this micromechanical picture, we are able to explore the role of inter-particle potential in the nonlinear rheology of colloidal suspensions. In the flowing charged colloids, the shear stress sustained by DCR is estimated as $\sigma_{DCR}=G^\prime \gamma_M$, where $G^\prime$ is the modulus of the local elasticity that is similar to the storage modulus given in Fig. 1(a) \cite{rogers1}. As shown in Fig. 1 (b), the microscopically determined stress $\sigma_{DCR}$ quantitatively agrees with the macroscopic shear stress for charged colloids $\sigma_{CC}$ ($\sigma_{CC}=\eta\dot\gamma$) determined from rheometry when $Pe\le1$ (or $\dot\gamma \le 10$ s$^{-1}$).  This observation clearly reveals that the extended-range inter-particle potential causes the high shear stress, or equivalently the viscosity, in the flow of charged colloids by establishing the local elasticity at low and moderate shear rates. At higher shear rates ($Pe\gg1$) \cite{brady0}, $\sigma_{DCR}$ exhibits considerable deviation from $\sigma_{CC}$, indicating that the hydrodynamic effect \cite{brady1} becomes prominent and the inter-particle electrostatic potential does not dictate the flow behavior of the charged colloids. In fact, as shown in Fig. 1 (b), $\sigma_{CC}$ and $\sigma_{HS}$ become comparable when $Pe\gg1$.  This observation, from a macroscopic measurement, also suggests that the DCR no longer plays a critical role in determining the nonlinear rheology at high shear rates. Lastly, it is worth noting while higher-order correlated collisions can produce entropy-driven elasticity at high volume fractions (\textit{ca.} 0.58) \cite{brader1,pusey1,pusey2}, such a mechanism is less important in the current colloidal suspensions, due to the low volume fraction of particles. 

Computational studies \cite{yama1,yama2} have established a characteristic length scale for quantifying the heterogeneity of local topological fluctuations in glassy liquids and correlated its evolution, which reflects the deformation inhomogeneity, with the shear thinning phenomenon.  Experimentally, we find that the size of the DCR decreases with increasing shear rate (Fig. 4(e)), implying the observed shear thinning behavior is accompanied by diminishing deformation heterogeneity.  This result is consistent with the previous computational investigations.

In conclusion, using SANS and rheometry, we identify a dynamically correlated region (DCR) in charged colloids under shear.  This DCR, spanning over the distance of a few particle diameters, is sustained by the extended-range inter-particle potential. The size of DCR displays positive correlation with the nonlinear rheological behavior and should be considered as a measure of heterogeneity in stress. Our results might shed light on the mechanism of nonlinear flow phenomena in highly-supercooled and glassy liquids.

This work was supported by the U.S. Department of Energy, Office of Science, Office of Basic Energy Sciences, Materials Sciences and Engineering Division. This Research at SNS of Oak Ridge National Laboratory was sponsored by the Scientific User Facilities Division, Office of Basic Energy Sciences, U.S. Department of Energy. The rheological characterization was carried out at the Center for Nanophase Materials Sciences, which is a DOE Office of Science User Facility. We acknowledge National Institute of Standards and Technology, U.S. Department of Commerce, in providing the neutron research facilities. Finally, we appreciate the D22 SANS beamtime from the Institut Laue-Langevin.

\end{document}